\documentclass[traditabstract, article, a4]{aa}
\usepackage{longtable}
\usepackage{times}                            %paquet de tipus de lletra
\usepackage{graphicx}
\usepackage{rotating}
\usepackage{multirow}
\usepackage{amssymb}
\usepackage{amsmath}
\usepackage{amstext}
\usepackage{listings}
\usepackage{lscape}
\usepackage{txfonts}
\usepackage{float}
\usepackage{natbib}
\bibpunct{(}{)}{;}{a}{}{,}
\usepackage{color}
\usepackage{longtable}

\newcommand{\BE}{\begin{equation}}
\newcommand{\EE}{\end{equation}}
\newcommand{\iras}{IRAS~16547$-$4247}
\begin{document}

\title{Studying the non-thermal lobes of \iras\ through a multi-wavelength approach}
%\subtitle{<your subtitle>}

\author{P.~Munar-Adrover\inst{\ref{inst1}}
\and V. Bosch-Ramon\inst{\ref{inst1}}
\and J. M.~Paredes\inst{\ref{inst1}}
\and K. Iwasawa\inst{\ref{inst2}}
}

\institute{
Departament d'Astronomia i Meteorologia, Institut de Ci\`encies del Cosmos (ICC), Universitat de Barcelona (IEEC-UB), Mart\'{\i} i Franqu\`es 1, 08028 Barcelona, Spain  \email{pmunar@am.ub.es}\label{inst1}
\and
ICREA and Institut de Ci\`encies del Cosmos (ICC), Universitat de Barcelona (IEEC-UB), Mart\'i i Franqu\`es 1, 08028 Barcelona, Spain \label{inst2}
}

\date{Received 25 May 2013 /
Accepted ...}

\abstract{
In the recent years massive protostars have been suggested to be high-energy emitters. Among the best candidates is IRAS~16547-4247, a protostar that presents a powerful outflow with clear signatures of interaction with its environment. This source has been revealed to be a potential high-energy source because it displays non-thermal radio emission of synchrotron origin, which is evidence of relativistic particles. To improve our understanding of IRAS~16547-4247 as a high-energy source, we analyzed \textit{XMM-Newton} archival data and found that \iras\ is a hard X-ray source. We discuss these results in the context of a refined one-zone model and previous radio observations. From our study we find that it may be difficult to explain the X-ray emission as non-thermal radiation  coming from the interaction region, but it might be produced by thermal Bremsstrahlung (plus photo-electric absorption) by a fast shock at the jet end. In the high-energy range, the source might be detectable by the present generation of Cherenkov telescopes, and may eventually be detected by \textit{Fermi} in the GeV range.
}


\maketitle

\section{Introduction}

Massive young stellar objects (MYSOs) have been suggested to be a new population of gamma-ray sources \citep{ara07,rome08,bosch10}, and several attempts to identify gamma-ray candidates have been carried out by various authors \citep{munar-adrover11, AraudoRodriguez12, marti13}. It is still unclear how massive stars form, although it is clear that outflows, in particular jets of high speed and high kinetic power, are involved in the process \citep{gara99, reip01, gara03}. 

The jets of MYSOs can propagate through the molecular cloud that hosts the protostar and even break its boundaries. At the jet-end point, strong shocks are expected to form that lead to non-thermal emission. Non-thermal synchrotron emission has been identified in a few sources, for instance \iras\ \citep{gara03}, Serpens \citep{rod89}, W3(OH) \citep{wilner99}, and HH~80-81 \citep{mar93, carrasco10}. It has been recently reported possible non-thermal X-ray emission arising from HH~80-81 lobe \citep{Lopez-Santiago2013}, being it the first time that non-thermal emission coming from the lobe of a MYSO is detected. Among these sources, IRAS~16547-4247 seems a particularly good candidate to produce high-energy emission \citep{ara07, bosch10}. This source is likely a young O-type protostar and has an associated highly collimated outflow. Located at a distance of $2.9\pm0.6$ kpc, this protostar was associated with a triple radio-continuum source consisting of a compact central object and two lobes located symmetrically from the central source in the northwest-southeast direction at a projected distance of 0.14 pc \citep{gara03}. The radio emission from the central object has a spectral index of $\alpha=$ 0.49 ($F \propto \nu^{\alpha}$), consistent with free-free emission from a thermal jet, and the north and south lobes have spectral indexes of $-$0.61 and $-$0.33, respectively, characteristic of non-thermal emission. Infrared (IR) data show that \iras\ is one of the most luminous ($\sim6.2\times10^4$L$_{\odot}$) protostellar objects associated with an outflow known so far. The high IR luminosity of \iras\ implies a photon field with an energy density $u_{ph} \sim 2\times10^{-9}$ erg cm$^{-3}$ in the region. The total jet luminosity of \iras\ has been estimated to be on the order of $10^{36}$ erg s$^{-1}$. The size of the core in which \iras\ is embedded is 0.38 pc and has a density of $n_c \simeq 5\times10^5$cm$^{-3}$ \citep{gara03}, giving a high hydrogen column density $N_H = 3.0\times10^{23}$ cm$^{-2}$. In X-rays, \iras\ was observed by \textit{XMM-Newton} in 2004 and only upper limits were obtained \citep{ara07, bosch10}, although preliminary results of a more detailed analysis have recently been published \citep{MunarAdrover12}.

In this work, we present the final results of a deeper analysis of \textit{XMM-Newton} archival data from 2004 and revisit the modeling at high energies, taking into account Coulombian losses that were not considered in previous modeling. The reliable X-ray detection and the impact of the Coulombian losses imposes strong constraints on the multi-wavelength modeling. The main conclusions of this work are the likely thermal Bremsstrahlung origin of the X-ray emission, and the strengthening of the prediction that gamma-ray radiation indeed takes place under reasonable physical conditions of the jet termination region.

\begin{figure*}
\resizebox{\hsize}{!}{\includegraphics{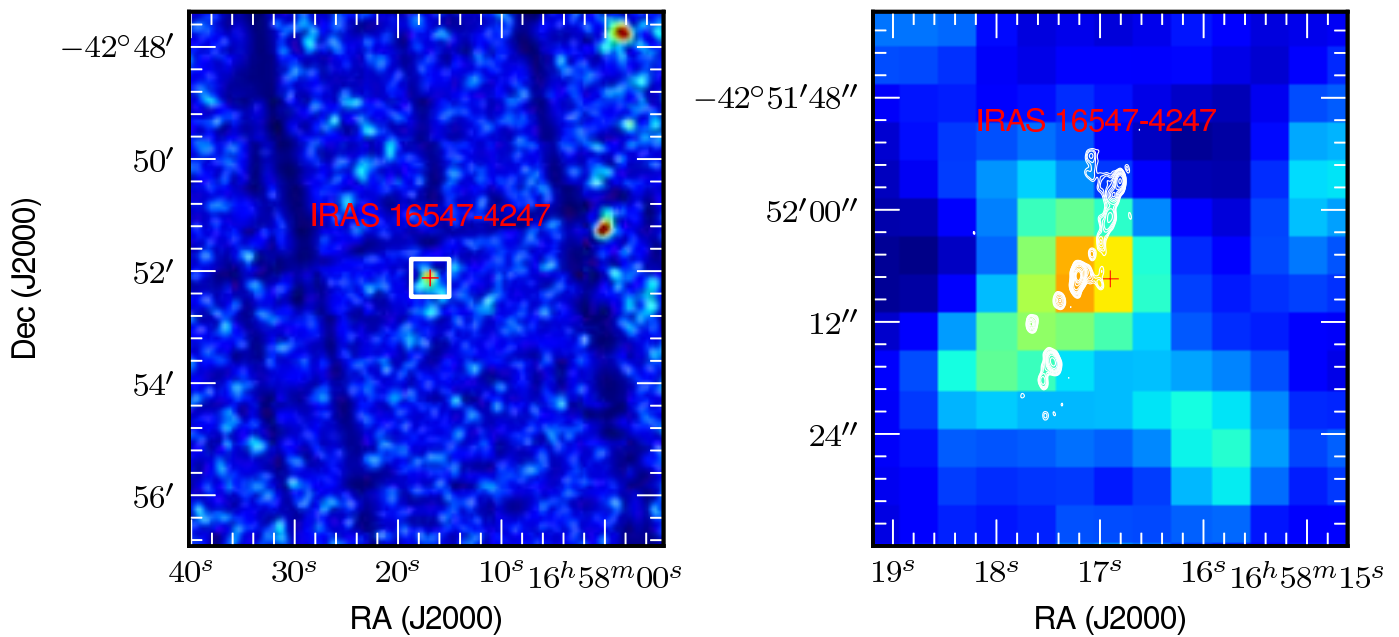}}
%    \put(-372,143.1){\color{white}\line(2,1){95}} 
%    \put(-372,128.5){\color{white}\line(2,-1){95}}
%    \put(-278.,190.2){\color{black}\line(2,1){79}}
%    \put(-278.,80.8){\color{black}\line(2,-1){79}}
  
  \caption{EPIC-pn camera image at the position of \iras\ in the 4.5$-$12.0 keV energy band. Left: wider field region to enable a comparison of the source with the surrounding background. Right: zoomed region of \iras\ (white box on left) overlapped with white radio contours \citep{rod08}. The red cross marks the nominal position of \iras\ . \label{Ximage}}
\end{figure*}

\section{Physical model for \iras}

\iras\ displays a thermal jet with non-thermal radio lobes at its termination region. The action of the jet onto the surrounding medium leads to two shocks: a forward shock (FS) that moves through the molecular cloud, and a reverse shock (RS) that moves inside the jet itself. These shocks can accelerate electrons and protons, which leads to the observed non-thermal radio spectrum through diffusive shock (Fermi I) acceleration (DSA) \citep{Drury83}.

%The Rankine-Hugoniot conditions for a strong adiabatic shock give us the relation between the densities and velocities downstream and upstream of the shock in its reference frame:
%\BE 
%\rho^{\prime} = 4~\rho
%\EE
%and 
%\BE
%v^{\prime} = \frac{v}{4}
%\EE
%where the variables with prime ($^{\prime}$) are the ones corresponding to the shocked region (downstream), and the ones without to the pre-shock region (upstream). 

We consider here the emission coming from the RS only, since the FS is expected to have a very low velocity and no significant energetics \citep{bosch10}. From now on, \textit{pre-shock} and \textit{post-shock} refer to the RS. We assumed that the mass density in the non-thermal emitter associated to the RS region is higher than the one given by the Rankine-Hugoniot conditions. This is a consequence of the mixing of RS and FS material at the contact discontinuity between the two shocked regions. This increase in density affects the intensity of the high-energy radiation produced in the RS non-thermal emitter, since the emission mechanisms relevant here depend strongly on the density, for instance relativistic Bremsstrahlung for electrons. 

The RS is expected to produce thermal and non-thermal emission through different processes. We used a one-zone model to compute the non-thermal emission of the RS, taking into account synchrotron, relativistic Bremsstrahlung, inverse Compton \citep{blumenthal-gould1970}, and $pp$-collisions \citep{kelner06}, as well as particle escape on a timescale $t_{esc}$, meaning that we solved the transport equation \citep{GinzburgSyrovatskii64} for a homogeneous emitter.
%\BE
%\frac{\partial N(t,E)}{\partial t}+\frac{\partial\left[b(E)N(t,E)\right]}{\partial E} + \frac{N(t,E)}{t_{esc}} = Q(E),
%\EE
%where $Q(E)$ is the energy distribution of injected particles and is assumed to be constant in time because the source lifetime is shorter than $t_{esc}$, and $b(E)$ includes all the cooling rates, $-E/t_{loss}$, which are relevant to our system. In this case, 
%The losses taken into account are synchrotron, relativistic Bremsstrahlung, inverse Compton losses for electrons , and $pp$ collisions for protons. 
We added to the physics of the model used by \cite{bosch10} by introducing Coulombian losses \citep{Lang99}
\BE
\begin{aligned}
\frac{dE_{Coul}}{dt} &= -\frac{2\pi e^4Z^2n_{c}}{m_ec^2}c \left[\ln\left(\frac{E}{m_ec^2}\right)-\ln\left(n_{c}\right)+\ln\left(\frac{{m_e}^3c^4}{\hbar^2e^2}\right)-1\right] \\
&\simeq  -5\times 10^{-19} n_{c}~{\rm erg~s^{-1}},
\end{aligned}
\EE
where E is the energy of the relativistic particles and $n_{c}$ is the medium number density. Introducing these losses has the effect of reducing the relativistic Bremsstrahlung luminosity at X-ray energies compared with previous results.

The parameters used in the model are summarized in Table \ref{tab_prop}, where they are listed in three different groups, depending on whether they come from the observational data or are required to explain the non-thermal or thermal emission.
%The source is embedded in a very dense cloud, with $n_c = 5\times10^5$ cm$^{-3}$ \citep{gara03} and located at a distance of 2.9 kpc \citep{rod08}. The energy density of the IR photon field is taken: $u_{ph} = 2\times10^{-9}$ erg cm$^{-3}$. The lobe size of \iras\ is fixed to $R_j = 1.6\times10^{16}$ cm, and the velocity of the jet to $v_j = 10^8$ cm s$^{-1}$ (\citealt{rod08}; 
The FS velocity is expected to be low, so the RS velocity is assumed to be equal to the jet velocity: $v_{RS} \sim v_j = 10^8$ cm s$^{-1}$. Non-thermal particles (electrons and protons) accelerated by the DSA mechanism have an uncooled energy distribution $Q(E) \propto E^{-\Gamma}$ with $\Gamma \sim 2$. The kinetic energy flux, in terms of luminosity (in erg s$^{-1}$), injected in the form of accelerated particles, electrons and protons separately, is taken as $10\%$\footnote{The adopted fiducial value of 10\% for the kinetic of the shock that enters non-thermal particles is similar to the efficiency of supernova-remnant shocks that accelerate cosmic rays (e.g. \cite{GinzburgSyrovatskii67}, for an early discussion).} of the RS luminosity: $L_{s} = \frac{1}{2}S_{s}\rho v_{RS}^3$, where $S_{s} = \pi {R_j}^2$ is the shock surface. For simplicity, we have adopted a model in which electrons and protons have the same injection luminosity. We note that with the same energy budget, explaining the radio data through synchrotron emission from $pp$-electron/positron secondaries in the proton-dominated case will require an increase in the magnetic field to explain the radio data with a somewhat lower gamma-ray flux for the same target density. More detailed studies of different proton-to-electron ratios can be found in \cite{ara07} and \cite{bosch10}.

We also considered that the RS can emit through thermal Bremsstrahlung. The temperature of the plasma in the RS is given by

\BE
\label{TShock}
T_{RS} = \frac{3}{32}\frac{m_{p}v_{RS}^2}{k_B} \simeq 1.1 v_{RS8}^2 ~{\rm keV}
\EE
for a fully ionized hydrogen plasma, where $k_B$ is the Boltzmann constant and $v_{RS8}=v_{RS}/10^{8}$cm s$^{-1}$. This temperature determines the maximum of the thermal emission from the RS. Given expected RS velocity conditions in IRAS~16547-4247, the peak of the thermal Bremsstrahlung emission is at $\sim$1~keV. 

The thermal Bremsstrahlung luminosity is given by
\BE
L_{Br} = \varepsilon_{Br}(n^2,T_{ps})\cdot V,
\EE
where $\varepsilon_{Br}$ is the Bremsstrahlung emissivity (in erg cm$^{-3}$ s$^{-1}$), which depends on the density of the targets (mainly ionized hydrogen nuclei) and on the post-shock temperature, $V = S_{s} X$ is the volume of the shocked region, and $X$ the depth of the shocked region. The value of $X$ can be estimated within a factor of a few using the following reasoning: the escape time of the region is
\BE
t_{esc} \sim \frac{R_{RS}}{v_{esc}},
\EE
where $v_{esc} \sim v_{RS}\sim v_j$.
The shocked material has a mass 
\BE
M_{s} = S_{s} \cdot \rho_s \cdot v_{RS} \cdot t_{esc},
\EE 
which can also be expressed as 
\BE
M_{s} = X \cdot S_{s} \cdot \rho_{s},
\EE
from where it is possible to determine $X$, which is $\sim R_{RS}$. If $X$ had a value $L_{Br} \gtrsim L_{s}$, the RS would be radiative, and the thermal Bremsstrahlung luminosity should be $L_{Br} = L_{s}$, since it cannot exceed the RS luminosity. Otherwise, if $X$ is such that $L_{Br} < L_{s}$, the shock is adiabatic, and the Bremsstrahlung luminosity will be roughly 
\BE
L_{Br}~=~\varepsilon_{Br}~\cdot~X~\cdot~S_{s} \sim \varepsilon_{Br}~\cdot~R_{RS}~\cdot~S_{s} .
\EE

As mentioned, we considered a mixing of the material of the FS region with material downstream the RS because of the formation of complex structures in the two-shock contact discontinuity \citep{Blondin1989}. This mixing increases the effective density downstream the RS at a certain distance from the shock itself through the presence of matter clumps from the FS region. Although a detailed treatment is beyond the scope of this work, note that for clump filling factors ($f$) and sizes ($R_c$) at least $f\cdot R_j > R_c$ (spherical clumps), the relativistic electrons/protons have high chances to enter these denser regions, and in general the populations lose most of their energy through relativistic Bremsstrahlung/$pp$-collisions. This is the reason why we adopted a density higher than the post-shock value for the non-thermal emitter. However, for the thermal emitter we adopted the post-shock density because this component of the RS material has enough temperature to emit X-rays, whereas in the denser clumps the temperature will be much lower. Given that the volume is mostly filled by normal shocked material, most of the synchrotron emission will come from this medium (assuming the same magnetic field).

\section{\textit{XMM-Newton} observations and analysis}

We analyzed archival data of the sky region surrounding \iras\ taken with the \textit{XMM-Newton}\footnote{The observation ID is 0200900101} X-ray telescope. Preliminary results of this analysis were published in \cite{MunarAdrover12}. The observation was carried out in pointing mode on 2004 September 24 and lasted for $\sim 29$ks. The medium-thickness optical blocking filter and full-frame mode were used in the three EPIC detectors (pn, MOS1, and MOS2) for the imaging observation. The data were analyzed using the \textit{XMM-Newton} Science Analysis System (SAS) version 12.0.1 and the set of {\tt ftools} from HEASARC. In a first step we cleaned the event files of the three EPIC detectors by removing the flaring high-background periods. For this purpose we selected the good time intervals (GTI) in which the count-rate for the most energetic events (E~$\geq$~10~keV) was below the standard threshold for each detector. After this cleaning process, the observation time that remained in each of the three detectors was $27.7$ks, $27.9$ks, and $19.9$ks for the MOS1, MOS2, and pn, respectively.\\ 

We searched for sources using the {\tt edetect-chain} command in SAS. This command concatenates a series of tasks that produce exposure maps, detector mask images, background maps, detected source lists and sensitivity maps. As a result, a list of detected sources is obtained containing count-rates, fluxes, and positions for each detected source, among other information. In this observation we detected a total of 22 X-ray sources. Most of them, including the brightest one, are unidentified sources. One source is coincident with a Be star, CD-42 11721.

The detection algorithm revealed for the EPIC-pn detector  a source coincident with the position of \iras\ with $\sim 34$ counts at a confidence level above 4$\sigma$, and a flux in the energy range 0.2-12.0 keV of $F_X = (3.79\pm0.96)\times 10^{-14}$ erg cm$^{-2}$ s$^{-1}$. Most of the flux ($\sim 85\%$) is detected in the 4.5-12.0 keV energy band. Figure~\ref{Ximage} shows the source detected by \textit{XMM-Newton} in this energy range and zooms in, overlapping the radio contours from \cite{rod08}. The peak of the detected X-ray emission seems to be coincident with the central radio source, although a visual inspection of the counts distribution might indicate a certain extension of the source. The analysis of the other images of the EPIC-pn detector shows no significant excess in the energy range below 4.5 keV at the nominal position of \iras . Lack of emission below 4.5 keV might be caused by strong photo-electric absorption in the dense environment in which the protostar is located. The source has not been detected in the EPIC MOS1 or in the MOS2 images with the {\tt edetect-chain} meta task, although hints above $2\sigma$ are obtained. An analysis of the images with the imaging software XIMAGE v4.5.1 ({\tt http://heasarc.gsfc.nasa.gov/xanadu/ximage/}) produced the results shown in Table \ref{table_stat}. In this table we list the instrument, the energy band, the counts of the source, and the intensity for each energy band, the signal-to-noise ratio (S/N), and a $3\sigma$ upper limit for all non-detection cases. The source is only detected in the highest energy band of the EPIC-pn detector.

\begin{table}[htb]
  \begin{center}
    \caption{Imaging analysis of \iras\ with XIMAGE. \label{table_stat}}        
    \scalebox{0.95}[0.95]{
    \begin{tabular}{ccr@{$\pm$}lr@{$\pm$}lcc}
    \hline
    \hline
EPIC   & E band & \multicolumn{2}{c}{Counts} & \multicolumn{2}{c}{Intensity}  & S/N & 3$\sigma$ U.L.\\
instr. & [keV]  & \multicolumn{2}{c}{}       & \multicolumn{2}{c}{[$\times10^{-4}$ s$^{-1}$]} &     &[$\times10^{-4}$ s$^{-1}$]               \\
\hline             
            & 0.2-0.5   &   6.7 & 7.3 &  5.0 & 5.4 &  0.9 & 24.0  \\
            & 0.5-0.1   &  -1.2 & 5.1 & -9.0 & 3.8 & -2.3 & 16.0  \\
pn          & 1.0-2.0   &   7.1 & 7.1 &  5.3 & 5.3 &  1.0 & 24.0  \\
            & 2.0-4.5   &  19.7 & 8.3 & 14.7 & 6.2 &  2.4 & 36.2  \\
            & 4.5-12.0  &  35.9 &10.0 & 26.8 & 7.8 &  3.4 &  $-$  \\
\hline
            & 0.2-0.5   & -11.1 & 1.8 & -4.0 & 0.7 & -6.2 &  5.7  \\
            & 0.5-0.1   &   3.7 & 5.3 &  1.3 & 1.9 &  0.7 &  8.5  \\
MOS1        & 1.0-2.0   &  13.3 & 8.2 &  4.8 & 3.0 &  1.6 & 15.0  \\
            & 2.0-4.5   &  13.9 & 8.1 &  5.2 & 2.9 &  1.7 & 15.1  \\
            & 4.5-12.0  &  19.5 & 9.3 &  7.0 & 3.4 &  2.1 & 18.4  \\
\hline
            & 0.2-0.5   &  -1.7 & 3.4 & -6.1 & 1.2 & -5.1 &  5.2  \\
            & 0.5-0.1   &  -3.4 & 4.4 & -1.2 & 1.6 & -7.8 &  6.1  \\
MOS2        & 1.0-2.0   &  -5.2 & 6.7 & -1.9 & 2.4 & -7.8 &  8.7  \\
            & 2.0-4.5   &  18.8 & 6.5 &  6.8 & 2.3 &  2.9 & 14.5  \\
            & 4.5-12.0  &  20.0 & 9.1 &  7.2 & 3.3 &  2.2 & 18.2  \\
\hline
     \end{tabular}}
  \end{center}
\end{table}

\begin{figure}
\resizebox{\hsize}{!}{\includegraphics[angle=270]{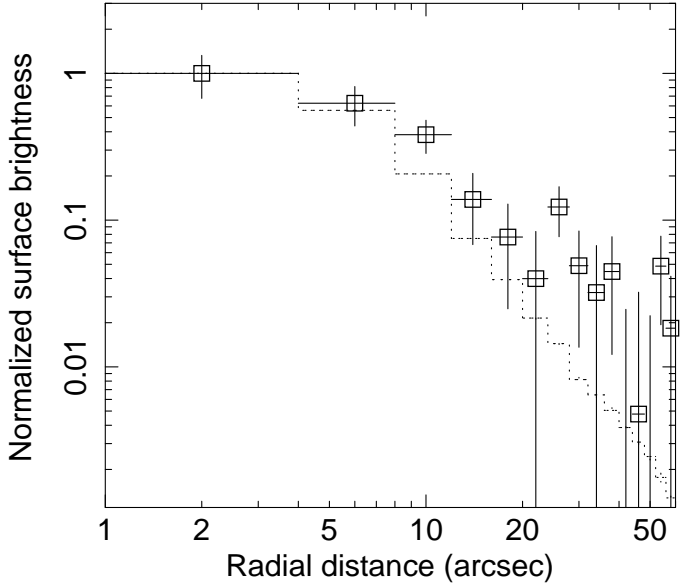}}
\caption{Radial profile adjusted to the image of the sum of the three EPIC detectors around the position of \iras\ (squared points) compared to a radial profile of a point-like AGN (dotted line). The PSF of the three EPIC instruments has been taken into account. \label{rad_prof}}
\end{figure}

To determine whether the X-ray emission is extended, as it is observed in the radio band, we adjusted a radial profile to the images of the three EPIC detectors in the 4.5-12 keV band (see Fig. \ref{rad_prof}) {\bf where our signal-to-noise ratio is maximum}. No clear extension can be inferred when adjusting a radial profile to the sum of the images of the three EPIC cameras or by comparing it with a point-like radial profile from an AGN, NGC 4395 (ObsID 0142830101, {\bf observed also on axis and at the same energy band as \iras}), although there are two points at radial distances of $\sim$10$^{\prime\prime}$ and $\sim$25$^{\prime\prime}$ that slightly depart from a point-like-source radial profile. However, the significance is not enough to claim a deviation from a point-like source. Spectral analysis is not possible because too few counts are detected by \textit{XMM-Newton}, but the detection above 4.5~keV shows that the emission is hard. A deeper observation with better angular resolution is required to elucidate the precise morphology and spectral characterization of IRAS~16547-4247 in X-rays.

\section{Model results and discussion}

\subsection{Non-thermal high-energy emission}

The model we used has an important difference compared with that used in \cite{bosch10}: we included Coulombian losses when computing the distribution of the accelerated particles for the non-thermal emission. This has an effect on the non-thermal radiation. In particular, the relativistic Bremsstrahlung emission is lower by almost one order of magnitude than the emission computed without Coulombian losses in the \textit{XMM-Newton} energy range. This makes the detected X-ray emission of \iras\ higher than the flux predicted by the non-thermal model. In the high-energy range (HE; 100 MeV$<$E$<$100 GeV) the model predicts a level of emission of about $10^{33}$ erg s$^{-1}$. The main contribution at high energies comes from relativistic Bremsstrahlung, although $pp$ interactions contribute slightly at GeV energies. The IC component is negligible at all energy ranges. The spectral energy distribution for one lobe from radio to gamma rays is presented in Fig.~\ref{model_all} as dashed lines.

The computed non-thermal emission at high energies for one lobe of the system is still relatively modest. However, the current and projected instruments sensitive to HE and very high-energy (VHE; E $>$ 100 GeV) will not be able to resolve whether the emission comes from one or the other lobe, or if it is the sum of both. For this reason we also computed the non-thermal and thermal emission from the southern lobe and added it to the computed emission of the northern lobe, obtaining the black solid line in Fig.~\ref{model_all}. This results in a slightly higher level of thermal emission at X-rays within the range of the model uncertainty, and a significant increase in the non-thermal emission at radio and gamma-rays. Accounting for both lobes, the source would be detectable by the current and next generation of Cherenkov telescopes for exposure times $\sim 50$ h, and even by \textit{Fermi} for exposure times of about six or more years.

\begin{table}[htb]
  \begin{center}
    \caption{\iras\ north observed properties and derived parameters. \label{tab_prop}}        
%    \scalebox{0.90}[0.90]{
    \begin{tabular}{llll}
    \hline
    \hline

\multicolumn{4}{c}{\iras} \\
\hline
         & L$_{\star}$ & [erg s$^{-1}$] & $5\times 10^{38}$ \\
         & d           & [kpc]          & 2.9 \\
Observed & n$_c$       & [cm$^{-3}$]    & $5\times 10^5$ \\
         &N$_H$       & [cm$^{-2}$]    & $3\times 10^{23}$ \\
         &$u_{IR}$    & [erg cm$^{-3}$]& $2 \times 10^{-9}$ \\
\hline
         &L$_j$       & [erg s$^{-1}$] & $1.5\times 10^{36}$ \\
%         &L$_{nt}$    & [erg s$^{-1}$] & $1.5\times 10^{35}$ \\
Non-thermal  &$\Gamma$    &                & $2.0$\\
omponent&R$_j$       & [cm]           & $1.6\times 10^{16}$ \\
         &v$_j$       & [cm s$^{-1}$]  & $10^8$ \\
%         &v$_{RS}$    & [cm s$^{-1}$]  & $10^8$ \\
         &B           & [G]            & $1.2\times 10^{-4}$ \\
\hline
Thermal  &T$_{RS}$    & [keV]          & $1.1$ \\
omponent&n$_s$       & [cm$^{-3}$]    & $5\times 10^3$ \\
         &X           & [cm]           & $1.6\times 10^{16}$ \\
    \hline
     \end{tabular}%}
  \end{center}
\end{table} 

\subsection{Connection with X-ray observations}

%The \textit{XMM-Newton} detection is at the higher energies at which the instrument is sensitive, between 4.5 and 12.0 keV. The inferred flux for the whole \textit{XMM-Newton} energy range is $F_X = (3.79\pm0.96)\times 10^{-14}$erg cm$^{-2}$ s$^{-1}$. 

\begin{figure*}
\resizebox{\hsize}{!}{\includegraphics{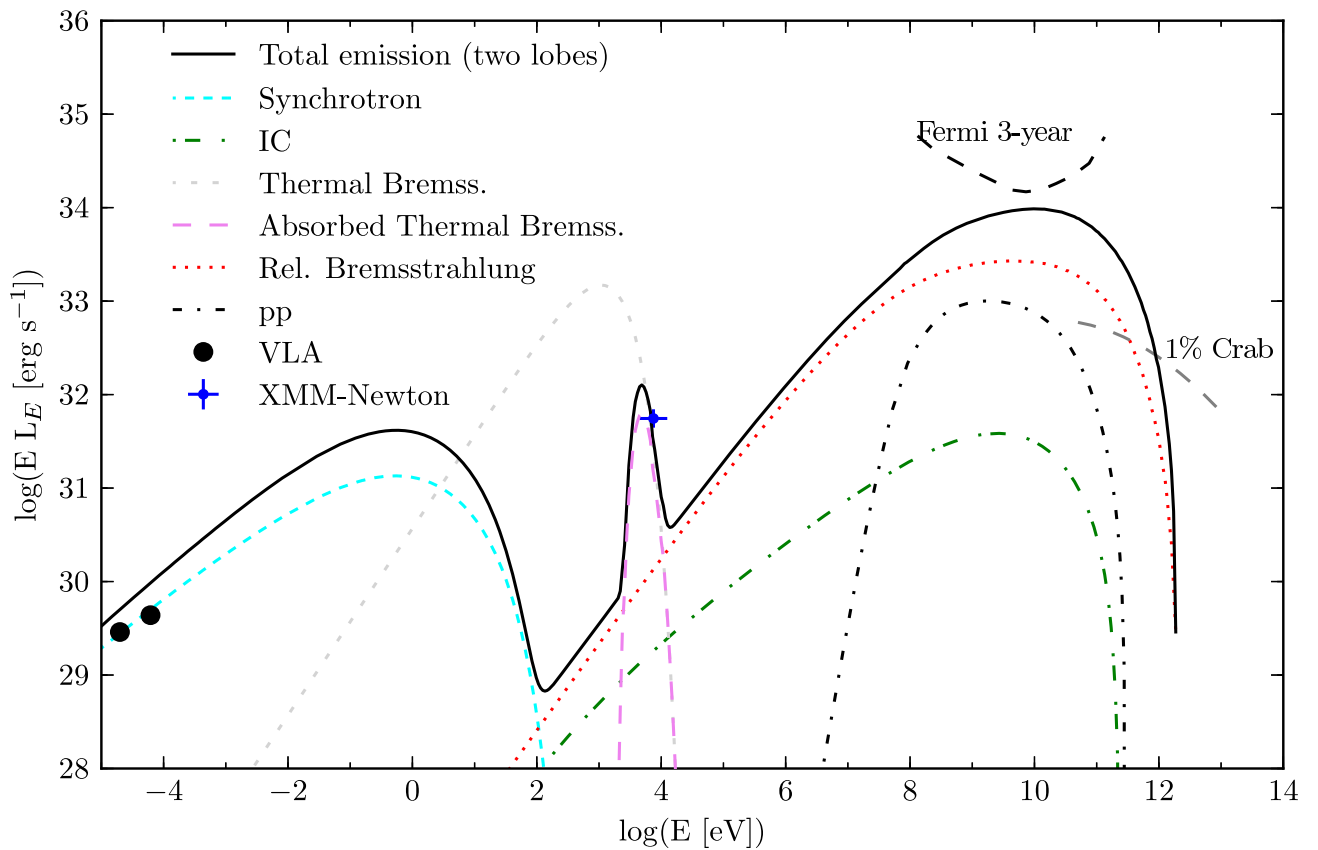}}
\caption{Computed non-thermal model including a thermal Bremsstrahlung absorbed component for the \iras\ north lobe. The black solid line represents the total thermal and non-thermal emission from the two lobes of the system. X-ray and radio data-points are also plotted, together with the \textit{Fermi-LAT} three-year (inner Galaxy) $5\sigma$ sensitivity curve and the SED for 1$\%$ of the Crab Nebula flux, the typical sensitivity of the current Cherenkov telescopes for $\sim$50 h of observation.}
\label{model_all}
\end{figure*}

The former attempt of fitting the X-ray data using only a non-thermal Bremsstrahlung component resulted in an unrealistic set of physical parameters \citep{MunarAdrover12}, which even predicted the source to be detectable by \textit{Fermi} in the second source catalog \citep{Nolan2012}. Therefore, to explain the X-ray detection, we introduced a thermal Bremsstrahlung component. Thermal X-rays are expected to originate from the RS region. With the known parameters of the source, one can calculate the post-shock RS temperature with Eq.~\ref{TShock}, which depends on the velocity of the RS: $v_{RS}\sim v_j = 10^8$ cm s$^{-1}$. The adopted density is that of the post-shock region, $n_{s} = 5\times10^3$ cm$^{-3}$ (different from the one in the non-thermal emitter, see Sect.~2). The temperature of the RS is $k_BT_{RS} \sim 1$ keV, which is approximately expected to be the peak energy of the thermal component. The resulting thermal Bremsstrahlung has a much higher flux than the detected X-rays and has its peak at lower energies (see the gray solid line in Fig.~\ref{model_all}). Since the detected emission can be affected by photo-electric absorption at low energies, as expected because the source is embedded in a molecular cloud, we introduced the photo-electric absorption effect \citep{MorrisonMcCammon1983}. The column density used to calculate the absorption is shown in Table \ref{tab_prop}, which is obtained from the value of the size and density from \cite{gara03}. This has the effect of dropping the flux of the thermal Bremsstrahlung emission and displaces the peak to higher energies, making it roughly consistent with the X-ray detection (pink dashed line in Fig.~\ref{model_all}).\\ \vspace{-0.1cm}

We note that the X-ray emission might also arise from other components of the system. Protostars have accretion disks that emit thermal radiation. The heated material can reach temperatures of up to T$\sim10^7$ K, and flaring events have been reported involving even higher temperatures, up to T$\sim10^8$ K \citep{Tsuboi98}. In low-mass protostars, models show that episodes of magnetic reconnection between the disk material and the protostar lead to X-ray emission and possibly even gamma-ray emission \citep{delValle2011}. The former might be also occuring in high-mass protostars \citep{pravdo09}. Thus, the detected X-ray radiation in \iras\ may come from the central protostar alone, from the RS, or be the sum of the contribution of the RS and the central source. Visual inspection of the source images from the EPIC-pn camera shows a distribution of counts that might indicate a certain extension. As noted, the adjusted radial profile to a point source is consistent with the source being point-like, but there are two points at radial distances of 10$^{\prime\prime}$ and 25$^{\prime\prime}$ from the centroid of the counts that slightly deviate from the point-like hypothesis. Unfortunately, the spatial resolution of \textit{XMM-Newton} does not allow us to discern wether the emission comes from the central source or from the lobe(s).

\subsection{Inferring medium conditions from radio data}
The observed radio spectrum is consistent with optically thin synchrotron emission. This indicates that the free-free opacity coefficient, $\tau_{ff}$, must be less than 1 in the region:
\BE
\tau_{ff} = \kappa_{ff} ~ l < 1,
\EE
where $\kappa_{ff}$ is the absorption coefficient \citep[see][]{Rybicki79} and $l$ is the size of the region. Using this constraint, we can derive an upper limit to the size of the region by estimating the opacity at the observed frequencies. To account for free-free absorption, we adopted a cloud density of $\sim 5\times 10^5$ cm$^{-3}$ \citep{gara03} and a temperature typical of an HII region, T$\sim10^4$ K, assuming full hydrogen ionization. Given the characteristics of the emission region, this upper limit results in $l \lesssim 10^{15}$ cm. This indicates that the region where free-free absorption is taking place is a thin shell surrounding the synchrotron emitter. The thinness of the absorbing region allows the non-thermal radio emission to escape without being significantly absorbed. This fact is consistent with that of strong photo-electric absorption of soft X-rays, as predicted by our model, because this implies that these photons are absorbed within a short distance, ionizing only a thin shell.

\begin{figure}
\includegraphics[width=0.49\textwidth, trim=-0.5cm 0cm 0cm 0cm]{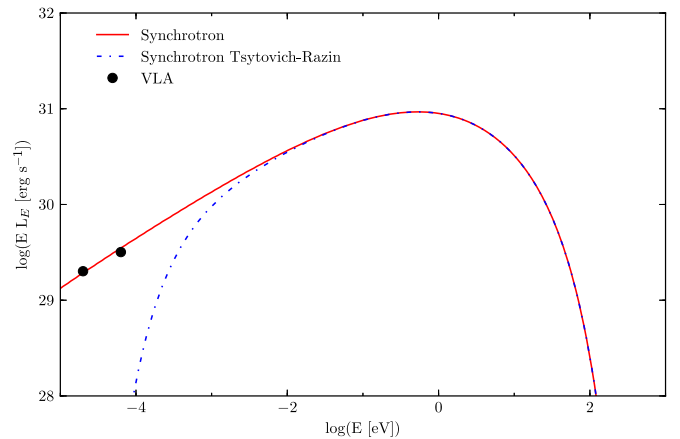}
\caption{Computed synchrotron spectral energy distribution and VLA data points. Two models are shown: one without absorption (except for synchrotron self-absorption) (red solid line) and one accounting for the Tsytovich-Razin effect (blue point-dashed line).}
\label{model_sync}
\end{figure}

The Tsytovich-Razin effect may be also important in the emitting region. This effect produces a suppression of the emission for frequencies below a characteristic cut-off frequency given by
\BE
\nu_R = 20 \frac{n_e}{B},
\EE
where $n_e$ is the number density of free thermal electrons and $B$ is the magnetic field strength \citep{Dougherty2003}. The descent in the level of the emission is due to a change in the refractive index of the medium where the radiation is produced. For the free  thermal electron density and magnetic field adopted in our calculations for the non-thermal emitter, $\nu_R \sim 8\times 10^{10}$~Hz~$\equiv 3.5\times 10^{-4}$~eV. However, the expected decrease in the radio spectrum due to the Tsytovich-Razin effect is not consistent with the VLA data, since the slope of the synchrotron spectral energy distribution (SED) at the detected frequencies is expected to be much harder than the one observed, and the intensity is expected to be lower (see Fig.~\ref{model_sync}). To be consistent, the density of the region would need to be at least 30 times less dense than the one we adoped in our model to account for non-thermal emission. However, the contact discontinuity between the jet and the cloud shocked material might be very complex. Mixing of the material of the RS with that from the FS is expected and the density excess probably takes the form of clumps. In this complex ambient, electrons can emit in radio through synchrotron radiation in a more diluted medium, and if they enter a clump of dense matter, they will also emit at high energies through relativistic Bremsstrahlung. The probability of entering one clump may be easily on the order of 1, as pointed out in Sect.~2, although we remark that most of the time particles are expected to be in the inter-clump medium.

%The time required for Rayleigh-Taylor instabilities to develop in the contact discontinuity is about the region size divided by the sound speed of the shocked material. Given that the shocked material in the forward shock is about 100 times denser than in the reverse shock, the sound speed will be 10 times smaller, and the mass injection rate will be about 10 times that of the reverse shock. However, in volume it would be just a 10$\%$ of the whole reverse shock region because of the higher density, which means that the probability of a relativistic particle to enter into a dense FS-region clump will be of order 0.1-1 depending on the denser clump size. This would imply that most of radio emission would come from the light regions of the reverse shock, with most of the radiated energy still channeled to the relativistic Bremsstrahlung channel in the dense clumps. The X-ray thermal emission of this dense clumps would be expected to fall in the UV range, because of the cooler temperature.
%
%The average density of the medium will still be roughly the one adopted in our calculations. This possibility also affects the free-free absorption, mentioned above in this section. If the RS accomplishes these conditions, with a density $n_{RS}\sim 1.7\times10^4$ cm$^{-3}$ (a factor 30 less dense than what we assume), the cut-off frequency of the Tsytovich-Razin effect goes to a lower value which makes the synchrotron SED consistent with the observed radio detections.

\section{Conclusions}

We have analyzed archival data of \iras\ observed by the \textit{XMM-Newton} X-ray Telescope. From our analysis we obtained a detection of the source above the 4$\sigma$ confidence level. Although the spatial resolution of the images is not enough to clarify whether the source is extended or not, the distribution of counts on the detector might indicate an extended nature, similar to the radio images. We were able to roughly explain the X-ray data by thermal Bremsstrahlung plus photo-electric absorption in the cloud. We compared our results and the VLA detections with a one-zone emission model, which improves previous efforts made to study and predict high-energy emission from this system \citep[e.g.][]{bosch10} by including Coulombian losses and the two non-thermal radio emitting regions in the calculations.  We accounted for mixing of the RS/FS material by adopting an average density $n_{RS}=5\times10^5$cm$^{-3}$ in the RS non-thermal emitter, which enhances the computed gamma-ray flux. Internally, however, there must be clumps of dense material that coexist with a (thermal X-ray emitting) medium at least $\sim$30 times less dense, which probably is the RS post-shock medium. This avoids a strong impact of the Tsytovich-Razin effect on the synchrotron emission at the radio band. Accounting for mixing of the RS-FS material in the non-thermal emitter and the contribution of the two non-thermal lobes, we predicted significant gamma-ray emission above the GeV energy band, which might be detectable by current and future HE and VHE gamma-ray telescopes.

\begin{acknowledgements} 
We acknowledge support by DGI of the Spanish Ministerio de Econom\'{\i}a y Competitividad (MINECO) under grant AYA2010-21782-C03-01. P.M-A acknowledges financial support from the Universitat de Barcelona through an APIF fellowship. V.B-R. acknowledges financial support from MINECO through a Ram\'on y Cajal fellowship and FPA2010-22056-C06-02. This research has been supported by the Marie Curie Career Integration Grant 321520. J.M.P. acknowledges financial support from ICREA Academia. 
\end{acknowledgements}

\bibliography{iras_16547-4247_2013}{}
\bibliographystyle{aa}

\end{document}